\documentclass[conference]{IEEEtran}
\IEEEoverridecommandlockouts
\usepackage{cite}
\usepackage{amsmath,amssymb,amsfonts}
\usepackage{graphicx}
\usepackage{textcomp}
\usepackage{xcolor}
\usepackage[capitalise]{cleveref}
\usepackage{multirow}
\usepackage{algorithm}
\usepackage[noend]{algorithmic}
\usepackage{siunitx}
\usepackage{subcaption} 

\def\BibTeX{{\rm B\kern-.05em{\sc i\kern-.025em b}\kern-.08em
    T\kern-.1667em\lower.7ex\hbox{E}\kern-.125emX}}
\begin{document}

\title{Adversarial Deep Feature Extraction Network for User Independent Human Activity Recognition}

% \author{\IEEEauthorblockN{Sungho Suh}
% \IEEEauthorblockA{\textit{German Research Center for Artificial Intelligence}\\
% Kaiserslautern, Germany \\
% sungho.suh@dfki.de}
% \and
% \IEEEauthorblockN{Vitor Fortes Rey}
% \IEEEauthorblockA{\textit{German Research Center for Artificial Intelligence}\\
% Kaiserslautern, Germany \\
% vitor.fortes_rey@dfki.de}
% \and
% \IEEEauthorblockN{Paul Lukowicz}
% \IEEEauthorblockA{\textit{German Research Center for Artificial Intelligence}\\
% Kaiserslautern, Germany \\
% paul.lukowicz@dfki.de}
% }

\author{\IEEEauthorblockN{Sungho Suh\IEEEauthorrefmark{1}\IEEEauthorrefmark{2}, Vitor Fortes Rey\IEEEauthorrefmark{1}\IEEEauthorrefmark{2} and Paul Lukowicz\IEEEauthorrefmark{1}\IEEEauthorrefmark{2}}
\IEEEauthorblockA{\IEEEauthorrefmark{1}German Research Center for Artificial Intelligence (DFKI), 67663 Kaiserslautern, Germany}
\IEEEauthorblockA{\IEEEauthorrefmark{2}Department of Computer Science, TU Kaiserslautern, 67663 Kaiserslautern, Germany}
\IEEEauthorblockA{Email: \{sungho.suh, vitor.fortes\_rey, paul.lukowicz\}@dfki.de}
}

\maketitle
% \IEEEpeerreviewmaketitle

\begin{abstract}
User dependence remains one of the most difficult general problems in Human Activity Recognition (HAR), in particular when using wearable sensors. This is due to the huge variability of the way different people execute even the simplest actions. In addition, detailed sensor fixtures and placement will be different for different people or even at different times for the same users. In theory, the problem can be solved by a large enough data set. However, recording data sets that capture the entire diversity of complex activity sets is seldom practicable. Instead, models are needed that focus on features that are invariant across users. To this end, we present an adversarial subject-independent feature extraction method with the maximum mean discrepancy (MMD) regularization for human activity recognition. The proposed model is capable of learning a subject-independent embedding feature representation from multiple subjects datasets and generalizing it to unseen target subjects. The proposed network is based on the adversarial encoder-decoder structure with the MMD realign the data distribution over multiple subjects. Experimental results show that the proposed method not only outperforms state-of-the-art methods over the four real-world datasets but also improves the subject generalization effectively. We evaluate the method on well-known public data sets showing that it significantly improves user-independent performance and reduces variance in results. 
\end{abstract}


\section{Introduction}
\label{sec:introduction}
	Human Activity Recognition (HAR) using wearable sensors is an important field in Ubiquitous computing where the task is to recognize which activity is being performed (sitting, walking, hammering, etc) given the data provided by the on-body sensors (IMU on the smartphone, watch, etc). This field spans more than two decades of research \cite{lara2012survey,wang2019deep}.
	%HAR methods range from classical ones to the more recent deep learning ones. In classical HAR methods, one selects or designs features that will be computed over sub-intervals (windows) of sensor data, using then a machine learning classifier, such as a random forest or a support vector machine (SVM) \cite{bao2004activity,chavarriaga2013opportunity,kwon2018adding}, to learn a mapping from features to activities. In the case of deep learning approaches, the raw (or lightly pre-processed) sensor data is fed into a deep neural network, which alone performs the feature extraction and classification.
	Recently, deep learning methods have achieved state-of-the-art performance in HAR by using convolutional neural networks (CNNs) as in \cite{yang2015deep} or by combining them with long short-term memory networks (LSTMs) as in \cite{ordonez2016deep}. Even more recently, transformer-like approaches have reached state-of-the-art results by employing self-attention \cite{mahmud2020human}.
	
	While much has been achieved, creating models that can generalize to unseen subjects is still a major challenge. As many studies have shown \cite{cutting1977recognizing, 7966182}, different people perform the same activities in different ways, which makes user recognition possible, but activity recognition more challenging. This is observed in practice by the gap in performance when evaluating by leaving out subjects instead of leaving out sessions.
	
	Approaches to deal with those limitations can be classified into classic approaches and deep learning-based methods. Classic approaches include directly selecting user-invariant classical features \cite{saputri2014user} or directly building one model per user \cite{7317739}. Building user models requires labeled data for all users, which is not realistic for many HAR applications as it may increase costs and deployment time. Selecting hand-crafted features is a possible solution, but may not be feasible as developing said features requires expertise in the target domain. Moreover, this approach may weaken the overall performance depending on the quality of the features available.
	
	Our work follows recent trends in deep learning approaches in this field: multi-task and adversarial learning. Recently, deep learning has been applied to this problem by exploring multi-task or even adversarial learning. In \cite{chen2020metier}, authors exploit user labels by combining HAR with subject identification in a multi-task learning framework that allows the model to focus on the relevant features for each user. This shows that taking user information into account can improve classification results, but it is not clear if models can learn to generalize beyond the available training subjects as they were not evaluated when leaving users out. Still, their work shows that deep learning models can clearly benefit from taking user information into account.  Other works such as \cite{sheng2020weakly} have shown that differences in the environment can, to some extent, be mitigated by employing similarity-based multi-task learning, but they have also not evaluated their model leaving users out and, moreover, their representation tends to create one cluster per subject with sub-clusters per activity, which may not favor generalization, being more similar to user-models.
	
	In the other direction, \cite{bai2020adversarial} use adversarial learning to generate a feature representation more robust regarding user variations. That is, instead of allowing the model to exploit user-specific information for classification, it should avoid leaking it as there is an adversary (a discriminator) whose objective is to separate subjects in the feature space. They achieve this by employing a Wasserstein Generative Adversarial Network (WGAN) \cite{arjovsky2017WGAN} and Siamese networks and showed that their method can generalize to new subjects without sacrificing performance, that is, still proving state of the art results. This may have other advantages besides performance, as neural networks are known to leak subject information \cite{iwasawa2017privacy} and applying adversarial learning can mitigate those concerns.
	
	While the deep learning-based methods such as \cite{bai2020adversarial} have achieved significant improvements, most of these methods still have limitations. Although adversarial learning improved the performance of the activity recognition by generalization, they cannot measure the degree of generalization of the latent features in the training procedure. In the adversarial learning procedure, feature representation modules as generators are trained to fool a discriminator, while the discriminator conducts binary classification to distinguish between two different features from representation modules with randomly chosen two subjects. Thus, it cannot measure the degree of generalization for all subjects and cannot generalize the feature representation over all subjects. Furthermore, their multi-view data representation module comprises three different networks to merge the sub-representations of different views. It requires large amounts of annotated data for human activity recognition to train this complex multi-view data representation module. However, it is challenging to collect a sufficiently large amount of datasets in personalized human activity recognition applications. Annotated medical datasets are limited due to the laborious labeling process, and sometimes legal issues associated with publicly sharing private personal information.
    
    To overcome these problems, we propose an adversarial subject independent feature extraction for human activity recognition. The proposed model is capable of learning a subject-independent embedding feature representation from multiple subjects datasets and generalizing it to unseen target subjects by using an adversarial learning procedure. The adversarial learning between a feature extractor and a discriminator, which distinguishes the extracted features from multiple subjects, learns the distributions of multiple subject data and extracts subject-invariant generalized features for activity recognition. To measure the degree of the feature generalization and align the distributions among the multiple subjects, we use the Maximum Mean Discrepancy (MMD) \cite{gretton2006kernel, long2015learning} regularization. The MMD regularization helps enhance the generalization ability of the proposed adversarial learning method.	Additionally, the proposed model adopts an encoder-decoder structure to preserve the characteristic of the original signals, which has been utilized in weakly supervised learning and feature representation \cite{zeng2017semi, varamin2018deep}. 
	
	The contributions of this paper can be summarized as follows. 
	\begin{itemize}
		\item A novel adversarial subject-independent feature extraction method is proposed for human activity recognition. We formulate the adversarial learning between feature extraction and subject discriminator and improve the performance of the activity recognition. 
		\item  We use the MMD regularization to enhance the generalization of the feature representation and measure the degree of the generalization.
		\item We design the feature extractor and reconstruction based on the encoder-decoder network structure to preserve the characteristic of the original input signals and extract important information for activity recognition.
		\item To validate the proposed method, we conducted experiments with four public real-world activity recognition datasets: Opportunity \cite{chavarriaga2013opportunity}, PAMAP2 \cite{reiss2012introducing}, MHEALTH \cite{banos2014mhealthdroid} and MoCapaci \cite{bello2021MoCapaci}. By the experiments on multiple datasets, we can verify the advantages and effectiveness of the proposed method for feature generalization and activity recognition.
	\end{itemize}
	
	The rest of the paper is organized as follows. Section \ref{sec:relatedwork} introduces the related works.
	Section \ref{sec:proposedmethod} provides the details of the proposed method. Section \ref{sec:experimentalresults} presents quantitative experimental results on the four datasets and analyzes the structure of the proposed model with ablation study. Finally, Section \ref{sec:conclusion} concludes the paper.

    \begin{figure*}[!t]
		\centering
		\includegraphics[width=0.85\linewidth]{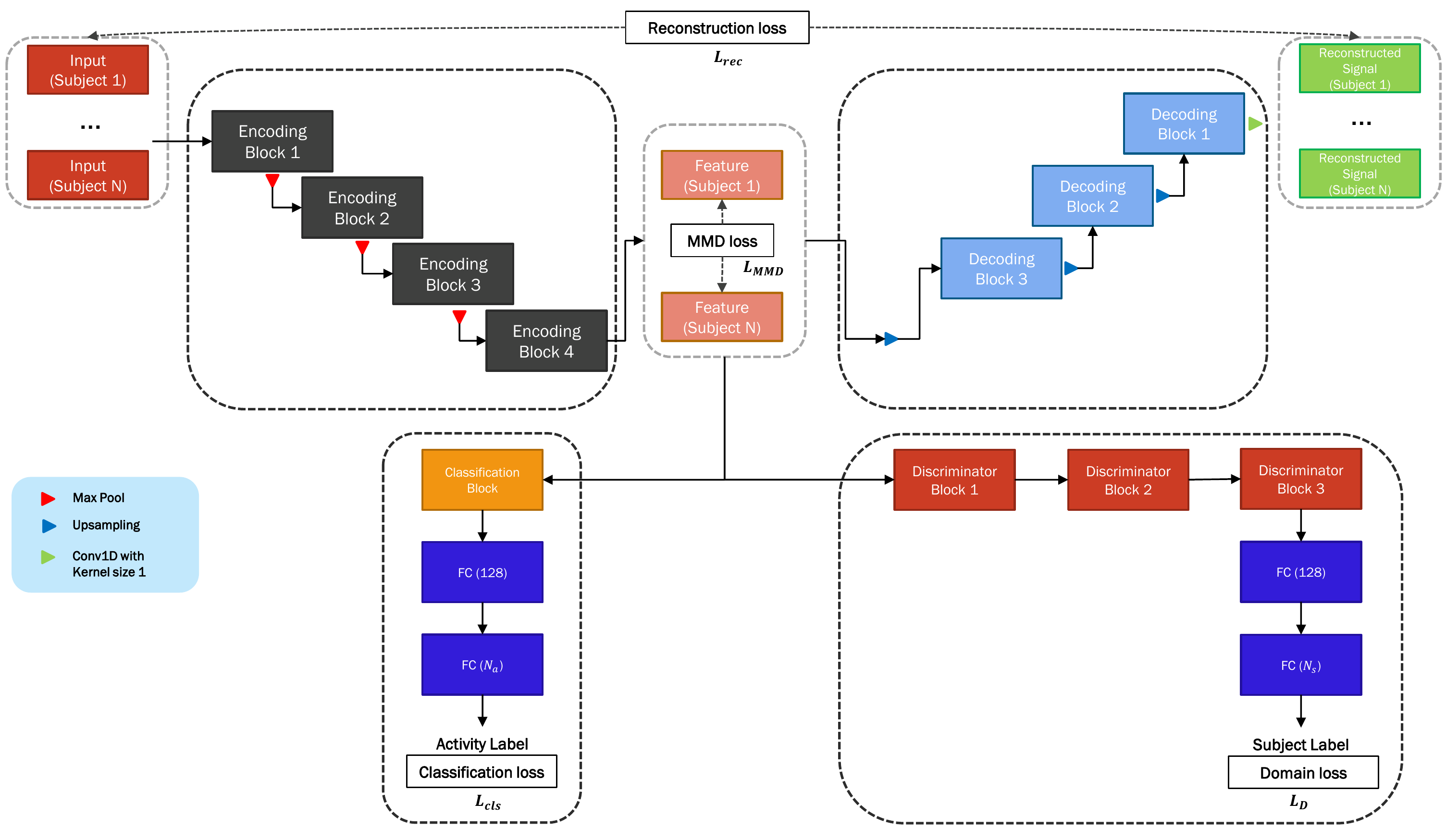}
		\caption{The overall network architecture of the proposed framework.}
		\label{fig:overview}
	\end{figure*}
	
\section{Related Work}\label{sec:relatedwork}

	Generally, there are two ways to capture the interpersonal variability: One is to increase the amount of training data from different subjects, and the another is to extract subject-independent features. The former is too expensive and impossible to collect and annotate data from different people. Recently, transfer learning methods have been investigated to solve cross-domain HAR problems. %, including cross-sensor-modalities \cite{morales2016deep}, cross-locations \cite{chiang2017feature}, and cross-subjects \cite{handiru2016optimized, zhao2020discriminant}. 
	Domain adaptation is the particular branch of transfer learning, that measures data distribution heterogeneity and aligns among data distributions \cite{cook2013transfer}. Deng et al. \cite{deng2014cross} proposed a cross-person activity recognition method using a reduced kernel extreme learning machine on the source domain, which classifies the target sample and the high confident samples and applies them to the training dataset. Zhao et al. \cite{zhao2011cross} introduced a transfer learning embedded decision tree algorithm that integrates a decision tree and the k-means clustering algorithm to recognize mobile phone-based different personalized activities by model adaptation. Wang et al. \cite{wang2018stratified} proposed a stratified transfer learning method that adopted the pseudo-leveling concept on the unlabeled target data by measure MMD between the feature spaces for the source domain data and the pseudo-labeled target data. Khan et al. \cite{khan2018scaling} proposed a heterogeneous deep convolutional neural network (HDCNN) that used a feature matching approach to adapt the pre-trained network, trained with the supervised source domain dataset, to an unlabeled target dataset collected from the smartwatch through the minimization of the discrepancy between two datasets after every convolutional and fully connected layer. They used Kullback-Leibler (KL) divergence as a distance measure between the pre-trained network and the target domain feature extractor network. Faridee et al. \cite{faridee2019augtoact} proposed an AugToAct framework which is directly aligned data distributions between source and target domains by combining augmentation transformations with deep semi-supervised learning to infer complex activities with the minimal labels in both source and target domains. AugToAct similarly performed domain adaptation as HDCNN but employed Jensen–Shannon (JS) divergence to minimize the discrepancy among two different domains instead of KL divergence. Akbari and Jafari \cite{akbari2019transferring} extracted stochastic features by training variational auto-encoder instead of deterministic feature extraction and employed the same network architecture of HDCNN to apply the feature matching approach to adapt the model to a target environment. Zhao et al. \cite{zhao2020local} a local domain adaptation method for cross-domain HAR, which aligns the distributions of source and target domains by using the MMD regularization. They first classified the activities into abstract clusters and mapped the original features into a low-dimensional subspace where the MMD between two clusters with the same label from different domains were minimized. The above domain adaptation approaches consider only a single source domain for domain adaptation task.
	
	Several studies have researched multiple source domain adaptation for sensor-based activity recognition. Some approaches focus on explicitly identifying the most relevant source domain among the multiple source domains with the target domain based on similarity measurement such as cosine similarity. Wang et al. \cite{wang2018deep} proposed a transfer neural network to perform knowledge transfer for activity recognition (TNNAR), which captures both the time and spatial relationship between activities. They considered explicitly selecting the most relevant domain from the multiple source domains based on the cosine similarity and applied the selected domain for domain adaptation to the target domain. Another group of approaches combines all the available source domains data and projects into lower-dimensional space, which is further processed by the classifier. Jeyakumar et al. \cite{jeyakumar2019sensehar} proposed a SenseHAR, which is a sensor fusion model for each device that mapped the raw sensor values to a shared low dimensional latent space. They mitigated the heterogeneous data distribution and assigned labels to the unlabeled data. However, the above-mentioned approaches did not consider multiple source domain data simultaneously, and the approaches cannot capture the uncertainty within the classification tasks.

	Recently, Multi-task and generative adversarial learning (GAN) \cite{goodfellow2014generative}-based methods have been introduced to solve the different data distribution problems. Chen et al. \cite{chen2020metier} proposed a deep multi-task learning-based activity and user recognition (METIER) model, which combines activity recognition and user recognition with a multi-task model. The model shares parameters between the activity recognition module and the user recognition module, and the activity recognition performance can be improved by the user recognition module employing a mutual attention mechanism. Sheng et al. \cite{sheng2020weakly} proposed a weakly supervised multi-task representation learning, which used Siamese networks to exploit a temporal convolutional network as a backbone model. Bai et al. \cite{bai2020adversarial} introduced a discriminative adversarial multi-view network, which extracts multi-view features from temporal, spatial, and Spatio-temporal views using CNN, and generalizes the multi-view features by employing WGAN and Siamese network architecture to decrease the variants between the extracted features from different subjects.

\section{Proposed Method}\label{sec:proposedmethod}
	
	\subsection{Problem Formulation}
	In this section, we introduce an adversarial subject-independent feature extraction method based on the encoder-decoder structure and GAN scheme for human activity recognition. Specifically, the task is to use data collected by sensor at different positions on the human body to predict one of $n_a$ activity labels. Let $X = [x_1, ..., x_n]$ be the time-series sensor data obtained by applying a sliding window of size $n_w$ of the $n_c$ available sensor channels  with $x_i \in \mathbb{R}^{n_c \times n_w} $ consisting of the sensor data for that window and $Y = [y_1, ..., y_n]$ the corresponding activity label set of $X$, and $S = [s_1, ..., s_{n}]$ the corresponding subject label set of $X$. We assume our training data $\{X_{src},Y_{src},S_{src}\}$ shares the same $n_a$ activities and sensor types with the test data $\{X_{tgt},Y_{tgt},S_{tgt}\}$ with unseen subjects in the training data.
	% 	Additionally, let $X_t = [x_{t,1}, ..., x_{t,n_w}]$ denote the unsupervised time-series sensor data of the target subject without the corresponding activity label and the corresponding subject label is $n_s+1$, where $n_s$ is the number of subjects in the source data. We assume that the source data contain the activity classes of the target data and the number of the activity classes denotes $n_a$.
	The goal of the proposed method is to generalize extracted features from the $S_{src}$ source subjects to the target subjects $S_{tgt}$ and improve the performance of the overall activity classification.

	\subsection{Network Architecture}
	% Overall proposed framework
	% Four independent network architectures with figures and tables
	The proposed neural network architecture is composed of four independent networks:
	\begin{enumerate}
		\item a feature extractor that acts as encoder of the encoder-decoder network structure and its role is to create our user-independent embedding features for activity recognition.
		\item a reconstructor that acts as a decoder in our encoder-decoder network structure and its role is to recompute the original signal from the embedding features so that they preserves the characteristics of the input signals.
		\item an activity classifier that predicts labels using the embedding features.
		\item a subject discriminator whose task is to distinguish the subject label from which the embedding features originated. It is used in an adversarial learning framework with the feature extractor, which tries to fool it by providing features that cannot be discriminated between subjects.
	\end{enumerate}
	\cref{fig:overview} presents the overall network architecture of the proposed framework with the four independent networks.
	
	% THe encoder-decoder structure
	%The encoder-decoder structure is generally used for unsupervised reconstruction-based methods.
	The encoder aims to map the input space $\mathcal{X}$ to a common embedding feature space $\mathcal{E}$, and the decoder reconstructs from the embedding feature space to its original input space. In this work, we denote the encoder as the feature extractor $Q: \mathcal{X} \rightarrow \mathcal{E}$ and the decoder as the reconstructor $P: \mathcal{E} \rightarrow \mathcal{X}$. The embedding feature space $\mathcal{E}$ from the feature extractor preserves the characteristics of the input signals so that the reconstructor can reconstruct the original signal. The feature extractor is composed of four encoding blocks, each including two convolution layers and a batch normalization \cite{ioffe2015batch} respectively, that extract increasingly abstract representation of the input signal and max-pool downsampling layers to encode the input signal into the embedding feature representations at multiple different levels. The structure of the reconstructor is similar to that of the feature extractor. The reconstructor is equipped with three decoding blocks, each including two convolution layers and a batch normalization respectively, and upsampling layers. 
	
	% Feature extractor and classifier
	The activity classifier $C$ predicts the activity labels using the embedding features $C: \mathcal{E} \rightarrow \mathcal{Y}$, where $\mathcal{Y}$ denotes the activity label space. The network of the activity classifier can be designed with long short-term memory (LSTM) \cite{hochreiter1997long}, ResNet50 \cite{he2016deep}, self-attention \cite{vaswani2017attention}, and so on. In this paper, the activity classifier is designed with a max-pool downsampling layer, a convolution layer, and two fully connected layers. The activity classifier's loss will also influence the feature extractor, as they are both trained to minimize the supervised classification. 
	% 	The activity classifier incorporates label information into training by the supervised learning setting and the embedding feature space should capture discriminative information on the activity label. Therefore, the feature extractor and activity classifier are trained to minimize the supervised classification loss $\mathbb{L}_{cls}$ simultaneously.
	
	% Feature extractor and subject discriminator
	The subject discriminator distinguishes the subject label from which the embedding feature. The goal of the proposed method is to extract subject-invariant features from multiple domains. In other words, the feature extractor generates a common embedding feature space for multiple subjects. Though the training procedure with the supervised activity classification encourages learning the data distribution of the activities, the extracted embedding features still contain subject-specific information. By adopting an adversarial manner between the subject discriminator and the feature extractor, the subject discriminator is trained to distinguish the subject label from the embedding features and the feature extractor is trained to fool the subject discriminator. A strong subject discriminator can train the feature extractor to generate embedding features that generalize across data from different subjects. The subject discriminator is composed of three discriminator blocks, each including a convolution layer, batch normalization, and a dropout respectively, and two fully connected layers. To align the distributions among the $N+1$ source and target subjects and further generalize the embedding feature representation, we use the MMD \cite{gretton2006kernel, long2015learning} regularization. %The detailed structure of the proposed adversarial subject-independent networks is shown in \cref{tab:networkarchitecture}.

	\subsection{MMD-Regularized Adversarial Learning for Multi-Subject Generalization}
	% Loss functions: MMD loss, reconstruction loss, classification loss, subject discriminator loss
	As shown in \cref{fig:overview}, we define four loss functions to train the four independent networks: 1) a reconstruction loss $\mathbb{L}_{rec}$, 2) a classification loss $\mathbb{L}_{cls}$, 3) a domain loss $\mathbb{L}_D$, and 4) a MMD loss $\mathbb{L}_{MMD}$. Let $P$, $Q$, $C$, and $D$ denote the reconstructor, the feature extractor, the activity classifier, and the subject discriminator, respectively. The reconstruction loss is defined to minimize the difference between the original signal and the reconstructed signal as follows.
	\begin{figure*}
		\centering
		\includegraphics[width=0.85\linewidth]{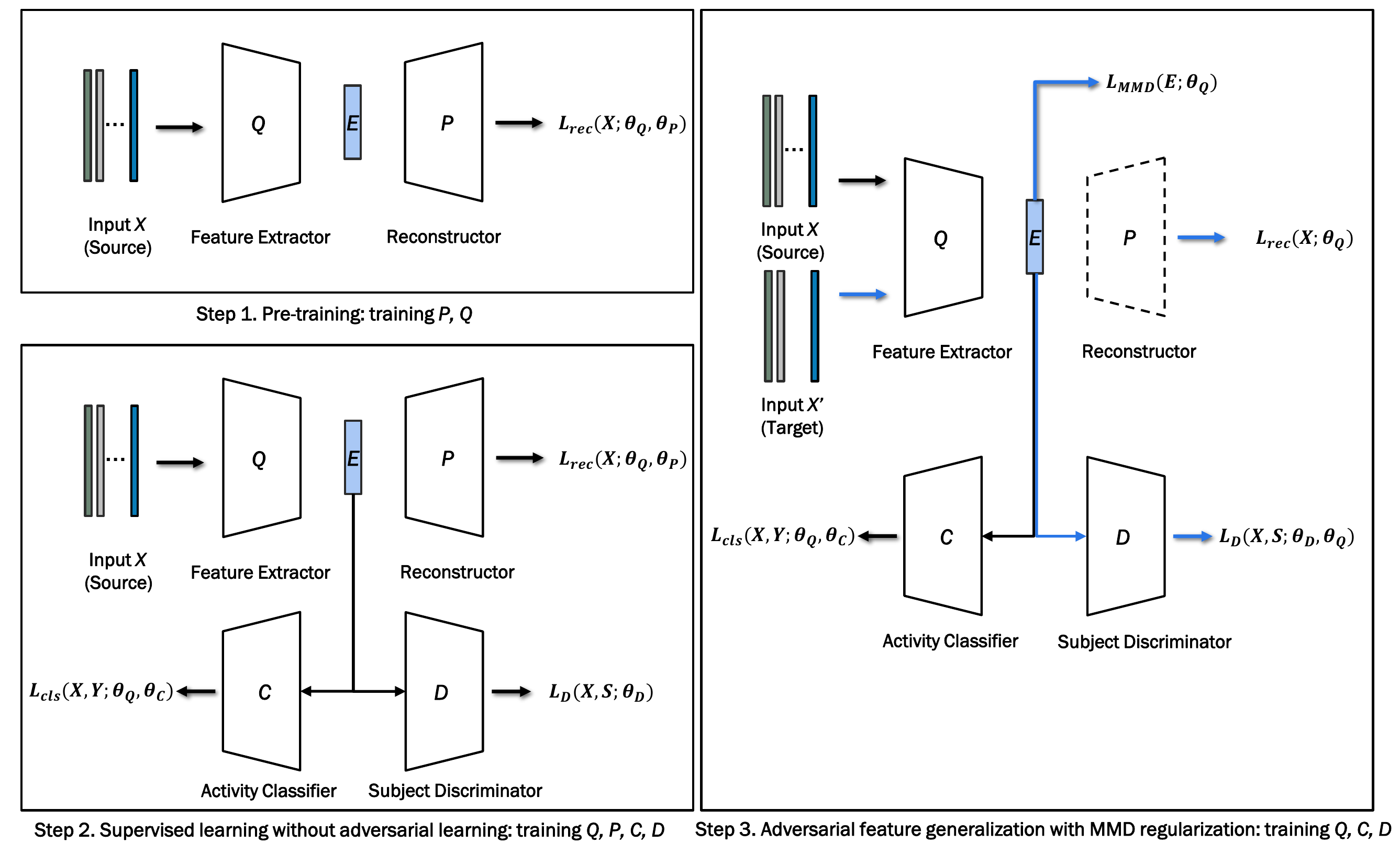}
		\caption{Training procedure, representing the steps described in \cref{subsec:training}. Solid lines indicate that the network is being trained, dashed lines indicate that the parameters of the network are fixed.}
		\label{fig:procedure}
	\end{figure*}
	\begin{equation}
		\label{eq:recon}
		\begin{split}
			\mathbb{L}_{rec} (X) =  \lVert P(Q(X)) - X \rVert_2^2,
		\end{split}
	\end{equation}
	where $X$ denotes the original sensor data and $P(Q(X))$ denotes the reconstructed signal. 
	
	In the proposed network architecture, the feature extractor and activity classifier are trained by supervised learning with given activity labels. The classification loss for the feature extractor and activity classifier is expressed as follows:
	\begin{equation}
		\label{eq:classification}
		\begin{split}
			\mathbb{L}_{cls} (X,Y) =  - \sum_{i=1}^{n} y_i \log C(Q(x_i)),
		\end{split}
	\end{equation}
	which is the cross-entropy classification loss.
	
	The goal of the subject discriminator is to distinguish which subject generated the embedding features. The subject discriminator is thus trained to minimize the domain loss:
	\begin{equation}
		\label{eq:domain}
		\begin{split}
			\mathbb{L}_D (X,S) =  -\sum_{i=1}^{n} s_i \log D(Q(x_i)),
		\end{split}
	\end{equation}
	
	Here, we address the MMD regularization to align the distributions among different subjects and to further improve the generalization of embedding features extracted by the feature extractor. The MMD is one of the most commonly used non-parametric methods to measure the distance of the distribution between two different domain datasets. The feature extractor represents the embedding features from the input signals $Q: \mathcal{X} \rightarrow \mathcal{E}$, and we let $Q(X_s) = E_s=[e_{s,1}, e_{s,2},...,e_{s,M}]$ and $Q(X_t) = E_t=[e_{t,1},e_{t,2},...,e_{t,N}]$ represent the embedding feature sets of two different subjects (domains). A mapping operation $\phi(\cdot)$ projects the features of two different domains onto the reproducing kernel Hilbert space (RKHS) $\mathcal{H}$ \cite{gretton2006kernel}, and calculates the mean distance between the two domains in RKHS. The MMD between two subjects can be calculated by using the function $\phi(\cdot)$ as follows:
	\begin{equation}
		\label{eq:MMD2}
% 		\begin{split}
			MMD(E_s, E_t)^2 = \left\lVert \frac{1}{M} \sum_{i=1}^{M} \phi(e_{s,i}) - \frac{1}{N} \sum_{j=1}^{N} \phi(e_{t,j}) \right\rVert_{\mathcal{H}}^2
% 		\end{split}
	\end{equation}
	The key to calculate the MMD is to find the appropriate $\phi(\cdot)$ as a mapping function to map the two domains to RKHS $\mathcal{H}$. Thus, the mean difference between the two data distributions after the mapping is calculated as their difference. $\phi(\cdot)$ represents a characteristic kernel function as $k(e_{s,i}, e_{t,j}) = \langle \phi(e_{s,i}), \phi(e_{t,j})\rangle$, and the MMD can be rewritten as follows:
	\begin{equation}
		\label{eq:MMD3}
		\begin{split}
			MMD&(E_s, E_t)^2 = \frac{1}{M} \sum_{i=1}^{M} \sum_{j}^{M} k(e_{s,i},e_{s,i}) \\
			                  &+ \frac{1}{N} \sum_{i=1}^{N} \sum_{j}^{N} k(e_{t,i},e_{t,i}) 
			                  - \frac{2}{MN} \sum_{i=1}^{M} \sum_{j=1}^{N} k(e_{s,i}, e_{t,j})
		\end{split}
	\end{equation}
	Generally, the Gaussian kernel function $k(e_{s,i}, e_{t,j}) = \exp(-\frac{\sigma}{\lVert e_{s,i} - e_{t,j} \rVert^2 })$ is used as the kernel function in the MMD algorithm, which maps data to infinite-dimensional space. This MMD method is based on a single kernel transformation. In this work, we adopt the multi-kernel MMD (MK-MMD) \cite{long2015learning}, which is an extension of the MMD and the optimal kernel can be obtained by linear combination of multiple kernels. The kernel function is defined as the convex combination of $m$ positive semi-definite (PSD) kernels ${k_u}$. The total kernel $k$ is defined as follows.
	\begin{equation}
		\label{eq:MMDkernel}
		\begin{split}
			K \triangleq \Big\{ k=\sum_{u=1}{m} \beta_u k_u : \sum_{u=1}^{m} \beta_u = 1, \beta_u \geq 0, \forall u \Big\}
		\end{split}
	\end{equation}
	where $k$ is weighted by different kernel $k_u$, $\{ \beta_u\}$ is the coefficient, which is the weight of $K$, to ensure that the generated multi-kernel $k$ is characteristic. 
	
	Unlike the general domain adaptation, the goal of the proposed method is to generalize the features from multiple subject domains from either only the $N$ source subjects or both $N+1$ source and target subjects. Thus, the overall MMD regularization loss $\mathbb{L}_{MMD}$ is described as follows.
	\begin{equation}
		\label{eq:overallMMD}
		\begin{split}
			\mathbb{L}_{MMD} (E_1,...,E_{N+1})	=  \frac{1}{(N+1)^2} \sum_{1 \leq i,j \leq N+1} MMD(E_i,E_j)
		\end{split}
	\end{equation}
% 	
	% Adversarial loss between feature extractor and subject discriminator
% 	
	The goal of the feature extractor is to extract the generalized feature among different subject domains, preserve the characteristics of the original data, and learn a class-discriminative feature representation. In other words, the feature extractor is trained to jointly minimize the losses of the reconstruction, classification, and MMD, and maximize the domain loss for the adversarial learning, simultaneously, whereas the reconstructor, the activity classifier, and the subject discriminator are trained to minimize the losses of the reconstruction, classification, and domain, respectively. Finally, The objective functions of the proposed method are defined as follows:
	
	% 	\begin{equation}
	% 		\label{eq:objective}
	% 		\begin{split}
	% 			\min_{C,Q,P} \max_{D} \mathbb{L}_{obj} (X,Y,S) = \lambda_{cls} \mathbb{L}_{cls} (X,Y)+ \lambda_{rec} \mathbb{L}_{rec} (X)+ \lambda_{MMD} \mathbb{L}_{MMD} (X) - \lambda_{D} \mathbb{L}_D (X,S)
	% 		\end{split}
	% 	\end{equation}
	\begin{equation}
		\label{eq:objective}
		\min_{C,Q,P} \max_{D} \mathbb{L}_{obj} (X,Y,S)
	\end{equation}
	where $\mathbb{L}_{obj}$ is
	\begin{equation}
	   \begin{split}
		    &\lambda_{cls} \mathbb{L}_{cls} (X,Y)+ \lambda_{rec} \mathbb{L}_{rec} (X) \\
		    &+ \lambda_{MMD} \mathbb{L}_{MMD} (X) - \lambda_{D} \mathbb{L}_D (X,S)
	    \end{split}
	\end{equation}

	where $\lambda_{cls}$, $\lambda_{rec}$ controls the relative importance of different loss terms, and $\lambda_{MMD}$ and $\lambda_{D}$ are hyperparameters that control the effect of domain generalization.     
	In summary, the reconstruction loss $\mathbb{L}_{rec}$ is computed to preserve the content information of the embedding features, the classification loss $\mathbb{L}_{cls}$ is used to improve the performance of the activity recognition, the MMD regularization term $\mathbb{L}_{MMD}$ helps to measure and align the distribution distance among different subjects, and the domain loss $\mathbb{L}_{D}$ hinders extracting subject domain-specific information from the feature extractor by the adversarial learning between the feature extractor and subject discriminator.
	
	\subsection{Training Procedure}
	\label{subsec:training}
	% Figure and Algorithm
	
	\begin{algorithm}
		\caption{Training procedure for adversarial generalized feature extraction using MMD regularization. We use default values of $\lambda_{cls}=5$, $\lambda_{rec}=5$, $\lambda_{MMD}=1$, $\lambda_D=1$}\label{Algo}
		\begin{algorithmic}[1]
			\REQUIRE Batch size $m$, Adam hyperparameters $\eta$, hyperparameter for $\lambda_{rec}$, $\lambda_{cls}$, $\lambda_{D}$, $\lambda_{MMD}$.
			\STATE \textbf{Input:} $X$, $Y$, $S$, $X_t$, $Y_t$, $S_t$
			\FOR{number of training iterations for step 1}
			\STATE Sample a batch $x$ from the training dataset $X$.
% 			\STATE Update reconstructor $P$ and feature extractor $Q$ by descending the gradient of \cref{eq:recon}:
			\STATE $\theta_P \gets \theta_P - \eta_P \nabla_{\theta_P} \mathop{\mathbb{L}_{rec}}(x;\theta_P)$, 
			\STATE $\theta_Q \gets \theta_Q - \eta_Q \nabla_{\theta_Q} \mathop{\mathbb{L}_{rec}}(x;\theta_Q)$ \hfill$\triangleright$\cref{eq:recon}
			\ENDFOR
			\FOR{number of training interation for step 2}
			\STATE Sample a batch $(x,y,s)$ from the training dataset $X$, corresponding activity label $Y$, and domain label $S$.
% 			\STATE Update subject discriminator $D$ by descending the gradient of \cref{eq:domain} with $Q$:
			\STATE $\theta_D \gets \theta_D - \eta_D \nabla_{\theta_D} \mathop{\mathbb{L}_D}(x,s;\theta_D)$ \hfill$\triangleright$\cref{eq:domain}
% 			\STATE Update activity classifier $C$, feature extractor $Q$ by descending the gradient of \cref{eq:recon} and \cref{eq:classification}: $\mathbb{L}_{step2}(x,y)=\mathbb{L}_{rec}(X) + \mathbb{L}_{cls}(X,Y)$
			\STATE $\theta_P \gets \theta_P - \eta_P \nabla_{\theta_P} \mathop{\mathbb{L}_{step2}}(x;\theta_P)$, 
			\STATE $\theta_C \gets \theta_C - \eta_C \nabla_{\theta_C} \mathop{\mathbb{L}_{step2}}(x,y;\theta_C)$, 
			\STATE $\theta_Q \gets \theta_Q - \eta_Q \nabla_{\theta_Q} \mathop{\mathbb{L}_{step2}}(x,y;\theta_Q)$ \hfill $\triangleright$\cref{eq:recon,eq:classification}
			\ENDFOR
			\FOR{number of training iteration for step 3}
			\STATE Sample a batch $(x,y,s)$ from the training dataset $X$, corresponding activity label $Y$, and domain label $S$.
% 			\STATE Update subject discriminator $D$ by descending the gradient of \cref{eq:domain} with $Q$:
			\STATE $\theta_D \gets \theta_D - \eta_D \nabla_{\theta_D} \mathop{\mathbb{L}_D}(x,s;\theta_D)$ \hfill$\triangleright$\cref{eq:domain}
% 			\STATE Update activity classifier $C$, feature extractor $Q$ by descending the gradient of \cref{eq:objective}
			\STATE $\theta_C \gets \theta_C - \eta_C \nabla_{\theta_C} \mathop{\mathbb{L}_{obj}}(x,y;\theta_C)$, 
			\STATE $\theta_Q \gets \theta_Q - \eta_Q \nabla_{\theta_Q} \mathop{\mathbb{L}_{obj}}(x,y,s;\theta_Q)$ \hfill $\triangleright$\cref{eq:objective}
			\STATE Sample a batch $(x_t,s_t)$ from the target dataset $X_t$, and corresponding domain label $S_t$.
			\STATE $(x', s')  \gets (concat(x, x_t), concat(s, s_t))$
% 			\STATE Update subject discriminator $D$ by descending the gradient of \cref{eq:domain} with $Q$:
			\STATE $\theta_D \gets \theta_D - \eta_D \nabla_{\theta_D} \mathop{\mathbb{L}_D}(x',s';\theta_D)$ \hfill $\triangleright$\cref{eq:domain}
% 			\STATE Update feature extractor $Q$ by descending the gradient of \cref{eq:domain} and \cref{eq:overallMMD}: $\mathbb{L}_{tgt} (x',s') = \lambda_{MMD} \mathbb{L}_{MMD}(x') - \lambda_{D} \mathbb{L}_{D}(x',s')$
			\STATE $\theta_Q \gets \theta_Q - \eta_Q \nabla_{\theta_Q} \mathop{\mathbb{L}_{tgt}}(x',s';\theta_Q)$ \hfill $\triangleright$\cref{eq:domain,eq:overallMMD}
			\ENDFOR
		\end{algorithmic}
	\end{algorithm}
	
	The training procedure for the proposed adversarial subject-independent feature extraction method is based on three different steps in order to train the four independent neural networks to accomplish our intent. \cref{fig:procedure} presents the training procedure with the detailed steps. Before training all the proposed networks, a pre-training step is done where we train the feature extractor and reconstructor only. This step helps improve the stability of the training afterwards by learning the content information of the original input first. Second, we train all four networks by supervised learning without adversarial learning and MMD regularization. In this step, we jointly minimize the losses of reconstruction and classification to capture the data distribution of each activity with the source data and train the subject discriminator first to prevent making it too easy to fool the subject discriminator in the adversarial learning process afterward. Finally, we can train a subject-invariant feature extractor by adversarial learning and MMD regularization. In this step, we freeze the reconstructor so that the feature extractor can now solely focus on classification and generalization. The training details for the proposed adversarial feature extraction method are summarized in \cref{Algo}. 

\section{Experimental Results}\label{sec:experimentalresults}

	\begin{table*}
		\caption{Comparison results with the state-of-the-arts on OPPORTUNITY, PAMAP2, MHEALTH, and MoCapaci datasets. The numbers are expressed in percent and represented as $mean \pm std$.}
		\label{tab:comparisonresults}
		\centering
% 		\resizebox{\linewidth}{!}{
		\begin{tabular}{lllccc}
			\hline
			Dataset & Task	&Method 			& $acc$				& $F_w$				& $F_m$\\
			\hline
			\multirow{8}{*}{Opportunity} &	
			\multirow{4}{*}{Locomotion} 	
			& MC-CNN			& 69.19 $\pm$ 6.77	& 68.12 $\pm$ 6.65	& 69.37 $\pm$ 6.35\\
			& &DeepConvLSTM	    & 72.52 $\pm$ 5.34	& 77.62 $\pm$ 5.26	& 70.08 $\pm$ 10.49\\
			& &Transformer-like & 70.18 $\pm$ 15.44	& 68.77 $\pm$ 18.37	& 67.68 $\pm$ 18.41\\
			& &Proposed		& \textbf{76.72 $\pm$ 1.62} & \textbf{76.84 $\pm$ 1.45}	& \textbf{79.16 $\pm$ 2.18}\\
			\cline{2-5}
			&\multirow{4}{*}{Gestures} 
			&MC-CNN			& 72.80 $\pm$ 5.06	& 69.38 $\pm$ 4.36	& 28.94 $\pm$ 8.92\\
			& &DeepConvLSTM	& 68.27 $\pm$ 6.61	& 70.07 $\pm$ 5.00	& 37.29 $\pm$ 7.13\\
			& &Transformer-like & 77.62 $\pm$ 3.17	& 72.55 $\pm$ 5.44	& 33.77 $\pm$ 13.65\\
			& &Proposed		& \textbf{78.58 $\pm$ 2.35} & \textbf{79.52 $\pm$ 2.04}	& \textbf{56.28 $\pm$ 2.88}\\
			\hline
			\multirow{4}{*}{PAMAP2} 
			& &  MC-CNN			& 79.77 $\pm$ 14.55	& 79.16 $\pm$ 15.92	& 72.72 $\pm$ 15.32\\
			& & DeepConvLSTM	& 76.23 $\pm$ 16.34	& 75.21 $\pm$ 17.45	& 67.54 $\pm$ 16.00\\
			& & Transformer-like & 82.88 $\pm$ 14.82	& 82.37 $\pm$ 15.80	& 74.84 $\pm$ 15.70\\
			& & Proposed		& \textbf{85.69 $\pm$ 10.76} & \textbf{85.85 $\pm$ 11.26}	& \textbf{77.84 $\pm$ 11.69}\\
			\hline
			\multirow{4}{*}{MHEALTH} 
			& &  MC-CNN			& 89.69 $\pm$ 8.93	& 87.91 $\pm$ 10.41	& 87.42 $\pm$ 10.72\\
			& & DeepConvLSTM	& 89.24 $\pm$ 6.69	& 86.99 $\pm$ 8.41	& 87.17 $\pm$ 8.08\\
			& & Transformer-like & 87.44 $\pm$ 8.10	& 85.45 $\pm$ 9.06	& 85.13 $\pm$ 9.71\\
			& & Proposed		& \textbf{96.72 $\pm$ 3.61} & \textbf{96.37 $\pm$ 4.31}	& \textbf{96.47 $\pm$ 4.04}\\
			\hline
			\multirow{4}{*}{MoCapaci} 
			& & MC-CNN			& 85.42 $\pm$ 6.42 & - & -\\
			& & DeepConvLSTM	& 85.42 $\pm$ 5.84 & - & -	\\
			& & Transformer-like & 70.00 $\pm$ 11.73 & - & -\\
			& & Proposed		& \textbf{87.77 $\pm$ 4.65} & - & -\\
			\hline
		\end{tabular}
	\end{table*}

	\subsection{Datasets and Evaluation Metrics}
	To evaluate the effectiveness of the proposed method for human activity recognition, four types of popular datasets were used, Opportunity \cite{chavarriaga2013opportunity}, PAMAP2 \cite{reiss2012introducing}, MHEALTH \cite{banos2014mhealthdroid} and MoCapaci \cite{bello2021MoCapaci}, that contain continuous sensor data of various sensors and different human activities by different participants. %The detailed information of the four datasets is summarized in \cref{tab:dataset}. 
	Regarding preprocessing for each dataset:
		\begin{itemize}
		\item Opportunity: has three types of sensors: body-worn sensors with 145 channels, object sensors with 60 channels, and ambient sensors with 37 channels. In this paper, we selected a dimension of 113 channels taking into account only the body-worn sensors including the IMUs and accelerometers, following the setup of \cite{ordonez2016deep}. We preprocessed all channels of sensor data to fill in missing values using linear interpolation and to normalize the data values per channel to interval $[0, 1]$ with manually set minimum and maximum values per channel as in \cite{ordonez2016deep}. We used a sliding window size of 64 with a sliding step of 16, which is close to two seconds of the sliding window and a 0.5-second step size. We use two types of annotations from the dataset. One is modes of locomotion and postures, such as Stand, Walk, Sit, and Lie is annotated with five classes. Another is 18 mid-level gestures such as Open Door, Close Door, and Clean Table. 
		
		\item PAMAP2: has 52 channels, containing a channel of heart rate, 17 channels per IMU. The full IMU sensory data is composed of 6 channels of acceleration data, 3 channels of gyroscope data, 3 channels of magnetometer data, and 3 channels of orientation. In this work, we selected a total dimension of 36 channels by removing a channel of heart rate, a channel of temperature per IMU, 4 channels of orientation per IMU, since the orientation of IMUs is mentioned as invalid in the data collection. Additionally, we remove six activities classified "Optional" in the dataset and the ninth subject since the "Optional" activities were collected by only one subject and the ninth subject executed only one activity. Thus, a total of 12 activities named "protocol" in the dataset from 8 subjects are used in this work. We preprocessed all channels of selected sensor data to fill in NaN values using linear interpolation. All samples were normalized to zero mean and unit variance per user. The IMU data was collected under the sampling frequency of $\SI{100}{\hertz}$ and we used a sliding window length of 200 (2 seconds) with a sliding step of 50 (0.5 second).
		
		\item MHEALTH: provides a sampling rate for all sensing modalities of $\SI{50}{\hertz}$. To evaluate the proposed method with the iterative leave-one-subject-out cross-validation procedure, we augmented the dataset with a sliding window length of 200 (4 seconds) and a step size of 50 (1 second), unlike other methods \cite{nguyen2015recognizing, sheng2020weakly} used a sliding window length of 5 seconds and a step size of 2.5 seconds.
		
		\item MoCapaci: contains a total of 4 channels of capacitive data sampled at $\SI{100}{\hertz}$ per channel. The data of each gesture instance was filtered by a fourth-order Butterworth bandpass filtered from $\SI{1}{\hertz}$ to $\SI{10}{\hertz}$, and normalized by subtracting the average of the first and last values of the gesture. The length of the signal was resampled to 400 and we use the length of the signal as a sliding window length.
	\end{itemize}
	
	 To evaluate not only the performance of the proposed model but also how much performance varies depending on the subject, we adopt the leave-one-subject-out cross-validation procedure that all data from a subject is used as test set while all data from other subjects are used as a training dataset. The evaluation was repeated two times on each test set.

	To evaluate and compare the performance of the proposed method with others, we adopted three evaluation metrics, which are used in various human activity recognition studies\cite{ordonez2016deep, ma2019attnsense, bai2020adversarial}: accuracy $acc$, weighted F1-score $F_w$, and macro F1-score $F_m$.
% 	\begin{equation}
% 		\label{eq:accuracy}
% 		acc = \frac{TP}{TP + FN + FP + TN} 
% 	\end{equation}
% 	\begin{equation}
% 		\label{eq:weightedF1}
% 		F_w = \sum_{c=1}^{C} w_i \frac{2\times precision_c \times recall_c}{precision_c + recall_c}
% 	\end{equation}
% 	\begin{equation}
% 		\label{eq:MAPE}
% 		F_m = \frac{2}{C} \times \sum_{c=1}^{C} \frac{precision_c \times recall_c}{precision_c + recall_c}
% 	\end{equation}
% 	where $Recall=\frac{TP}{TP+FN}$, $Precision=\frac{TP}{TP+FP}$, and $TP$, $FP$, $TN$, and $FN$ denote the true positive, false positive, true negative ,and false negative values,	$C$ denotes the number of classes and $w_i=n_i/N$ is the proportion of samples of the $i$-th class with $n$ being the number of samples of the $i$-th class and $N$ being the total number of samples. 

    \subsection{Implementation Details}
	
	The experiments were all implemented using Python scripts in the PyTorch framework. Training procedures were conducted in the Linux system with four NVIDIA Quadro RTX 8000 GPUs. The hyperparameters for \cref{eq:overallMMD} were $\lambda_{rec}=5$, $\lambda_{cls}=5$, $\lambda_{MMD}=1$, and $\lambda_{D}=1$. Through various testing, it is observed that the parameters mentioned in the paper yield the best performance.
	%and the effect of the hyperparameter $\lambda_{MMD}$ for the MMD loss is addressed in \cref{subsubsec:hyperparameter}. 
	We chose the Adam optimizer \cite{kingma2014adam} with a learning rate of $\eta_C=\eta_Q=5 \times 10^{-5}$, $\eta_P=1 \times 10^{-4}$, $\eta_D= 1 \times 10^{-3}$, $\beta_1=0.9$, and $\beta_2=0.99$. The batch size was 500 for the Opportunity dataset, 200 for the PAMAP2 dataset, and 128 for the MoCapaci dataset, respectively. For the MK-MMD, the Gaussian kernel is applied in the MK-MMD, and its number is set to 5.

	\subsection{Comparison Results}
	% State-of-the-arts: MC-CNN [6], DeepConvLSTM [7], Transformer [8]
	The proposed method was evaluated on the Opportunity, PAMAP2, MHEALTH, and MoCapaci datasets. The three evaluation metrics were used to evaluate and compare the proposed method to deep-learning-based state-of-the-art methods including multi-channel time-series convolutional neural networks (MC-CNN) \cite{yang2015deep}, DeepConvLSTM \cite{ordonez2016deep}, and Transformer-like activity recognition method \cite{mahmud2020human}. MC-CNN is a CNN-based model consisting of three convolutional layers, two pooling layers, and two fully connected layers. DeepConvLSTM is a combined model of CNN and LSTM for activity recognition, that comprises four convolutional layers and two LSTM layers to learn both spatial and temporal correlations. The transformer-like activity recognition method introduced a self-attention mechanism to improve the performance of the human activity recognition based on wearable sensors. For the fair comparison study, we set up the same experimental conditions, such as splitting the training and test datasets, data preprocessing, the window size, and step size, as we addressed in the previous section.

	The Opportunity dataset is normally evaluated using two of its label types: the modes of locomotion recognition task and the gesture recognition task. We conducted the comparison experiment and introduce the experimental results on the two tasks separately. 
	
	\cref{tab:comparisonresults} shows the quantitative evaluation results on the Opportunity, PAMAP2, MHEALTH, and MoCapaci datasets. Because the Opportunity dataset for the gesture recognition has 18 types of activities and is severely imbalanced, the results in terms of $F_m$ are relatively lower and have a larger standard deviation than the results in terms of $acc$ and $F_w$. The results show that the proposed method achieves significantly higher performance on the Opportunity dataset for both locomotion and gesture recognition tasks in terms of all three of the measurements. Furthermore, the proposed method provides high-performance results with small inter-subject variation, whereas the state-of-the-arts give high standard deviation results relatively. Additionally, the performance variance on the PAMAP2 dataset is much larger than on the Opportunity dataset, because of the diversity between different subjects and the data quality. Nevertheless, the proposed method outperforms all the state-of-the-arts and achieves lowest standard deviation results among the state-of-the-arts. 
	
	The proposed method outperforms all the state-of-the-art methods on the MHEALTH datasets in terms of all three metrics. The proposed method achieves 7.03, 8.46, 9.05 percent points improvements over the best state-of-the-art method in terms of $acc$, $F_w$, and $F_m$, respectively. Additionally, the standard deviation of the performance by the proposed method is much smaller than that of the state-of-the-art methods. The comparison results demonstrate the superiority of the proposed model.
	
	Unlike other datasets, the MoCapaci dataset is perfectly class-balanced and subject-balanced. Thus, we evaluate the performance on the MoCapaci dataset only in terms of $acc$. The proposed method outperforms other methods and the variance is also smaller than the state-of-the-art methods.

	\begin{figure}
		\begin{subfigure}{\linewidth}
			\includegraphics[width=\linewidth]{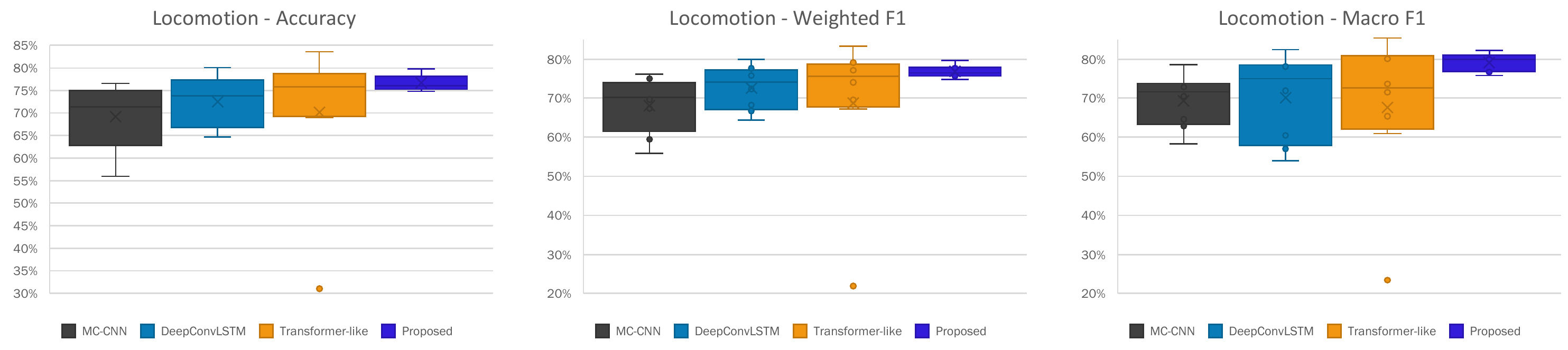}
			\caption{}
			\label{fig:comparisonlocomotion}
		\end{subfigure}%
		\hfill
		\begin{subfigure}{\linewidth}
			\includegraphics[width=\linewidth]{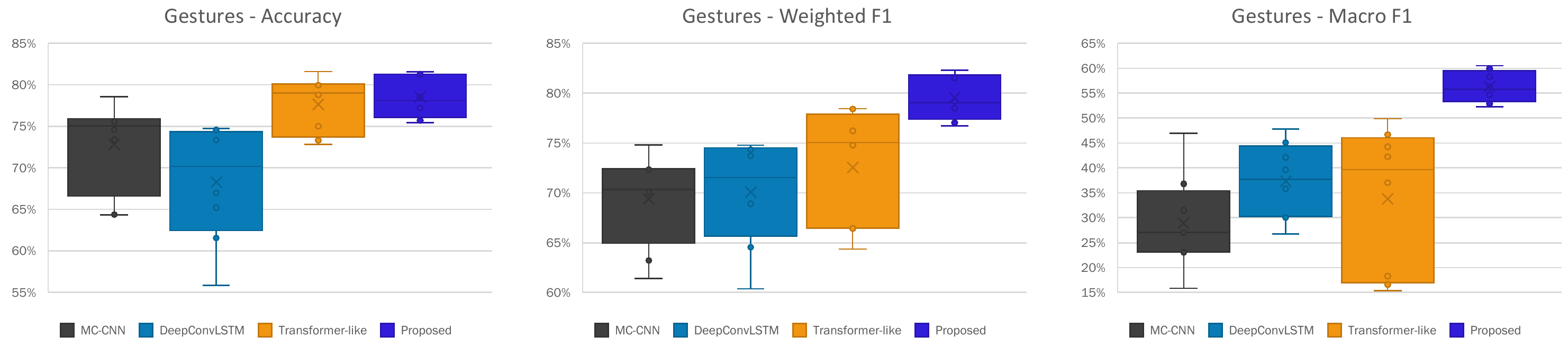}
			\caption{}
			\label{fig:comparisongestures}
		\end{subfigure}
		\begin{subfigure}{\linewidth}
			\includegraphics[width=\linewidth]{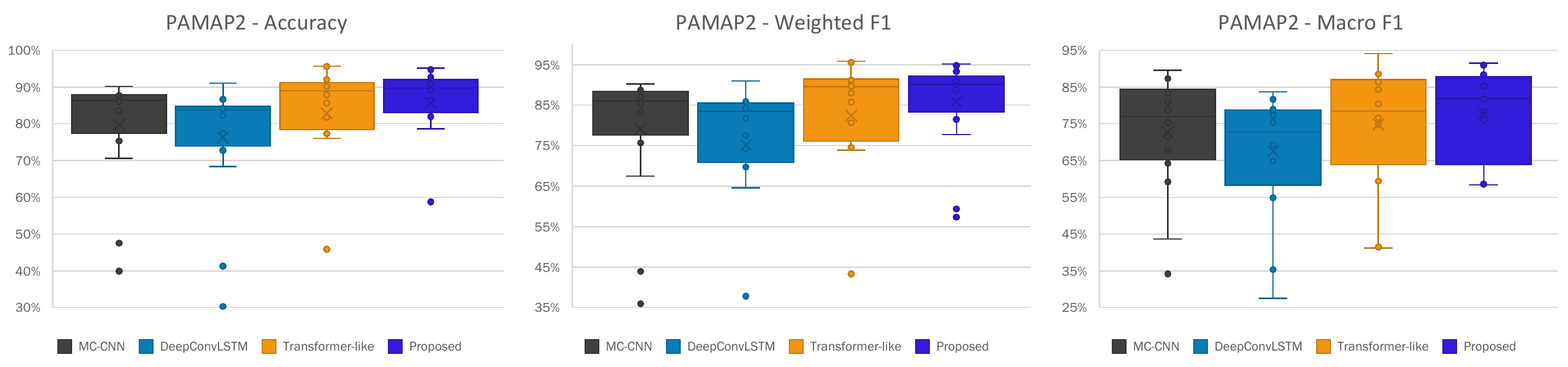}
			\caption{}
			\label{fig:comparison_PAMAP2}
		\end{subfigure}
		\begin{subfigure}{\linewidth}
			\includegraphics[width=\linewidth]{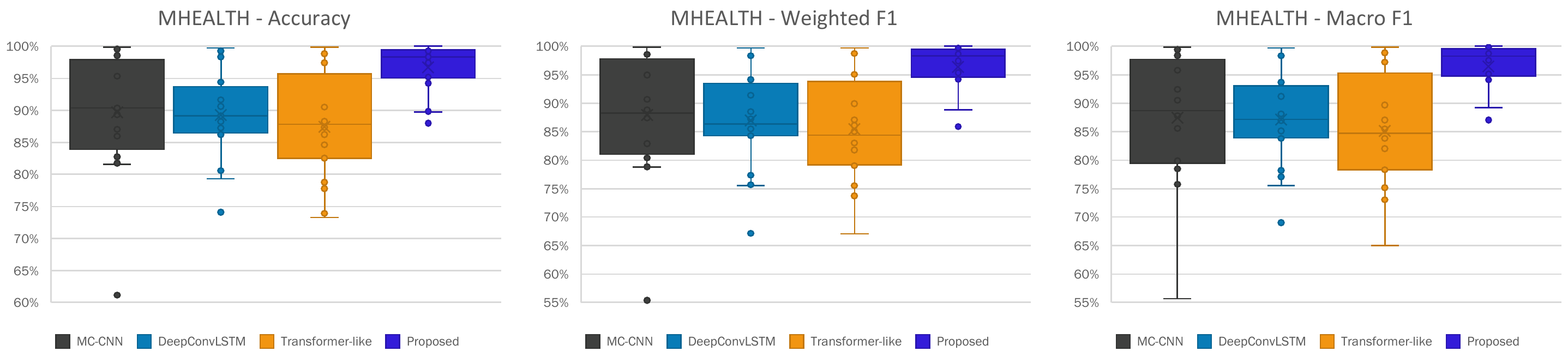}
			\caption{}
			\label{fig:comparison_MHEALTH}
		\end{subfigure}
		\caption{Box-and-whisker plots of comparison results with the state-of-the-art on OPPORTUNITY dataset (a) for the mode of locomotion recognition task, (b) for the gesture recognition, (c) PAMAP2 dataset, (d) MHEALTH dataset.}
		\label{fig:comparison_boxwhistker}
	\end{figure}
	
	\begin{table*}
		\caption{Evaluation results of ablation study for the proposed adversarial learning network architecture on OPPORTUNITY and PAMAP2 datasets. The numbers are represented as $mean \pm std$.}
		\label{tab:ablationstudy}
		\centering
% 		\resizebox{\linewidth}{!}{
		\begin{tabular}{lllccc}
			\hline
			Dataset & Task	&Method 			& $acc$				& $F_w$				& $F_m$\\
			\hline
			\multirow{8}{*}{Opportunity} &	
			\multirow{4}{*}{Locomotion} 	
			&  NoAdv			& 72.08 $\pm$ 3.30	& 72.26 $\pm$ 2.99	& 72.03 $\pm$ 6.43\\
			& &OnlySupervised	& 74.05 $\pm$ 2.66	& 74.26 $\pm$ 2.25	& 76.40 $\pm$ 2.72\\
			& &NoMMD			& 76.15 $\pm$ 4.11	& 76.45 $\pm$ 3.52	& 77.74 $\pm$ 5.99\\
			& &1Stage			& 74.32 $\pm$ 1.92	& 74.41 $\pm$ 2.22	& 75.52 $\pm$ 2.99\\
			& &Proposed		& \textbf{76.72 $\pm$ 1.62} & \textbf{76.84 $\pm$ 1.45}	& \textbf{79.16 $\pm$ 2.18}\\
			\cline{2-6}
			&\multirow{4}{*}{Gestures} 
			&  NoAdv			& 73.28 $\pm$ 2.28	& 74.20 $\pm$ 1.98	& 43.67 $\pm$ 4.29\\
			& &OnlySupervised	& 77.25 $\pm$ 2.31	& 77.30 $\pm$ 2.18	& 49.05 $\pm$ 6.16\\
			& &NoMMD			& 76.55 $\pm$ 2.10	& 77.39 $\pm$ 1.91	& 50.28 $\pm$ 3.65\\
			& &1Stage			& 76.31 $\pm$ 2.75	& 77.13 $\pm$ 2.52	& 50.93 $\pm$ 4.27\\
			& &Proposed		& \textbf{78.58 $\pm$ 2.35} & \textbf{79.52 $\pm$ 2.04}	& \textbf{56.28 $\pm$ 2.88}\\
			\hline
			\multirow{4}{*}{PAMAP2} 
			& &  NoAdv			& 81.92 $\pm$ 14.55		& 81.99 $\pm$ 15.34		& 74.14 $\pm$ 15.35\\
			& &OnlySupervised	& 83.48 $\pm$ 11.36		& 83.60 $\pm$ 11.90		& 76.48 $\pm$ 13.33\\
			& &NoMMD			& 84.46 $\pm$ 13.86		& 84.61 $\pm$ 14.44		& 76.51 $\pm$ 14.51\\
			& &1Stage			& 85.27 $\pm$ 9.48		& 85.75 $\pm$ 9.74		& \textbf{77.92 $\pm$ 12.05}\\
			& & Proposed		& \textbf{85.69 $\pm$ 10.76}& \textbf{85.85 $\pm$ 11.26}	& 77.84 $\pm$ 11.69 \\
			\hline
		\end{tabular}
	\end{table*}
	
	In \cref{fig:comparison_boxwhistker}, we analyze the results of the proposed method and the stat-of-the-arts in the box-and-whisker plots in terms of $acc$, $F_w$, and $F_m$. The proposed method not only provides significantly better performance but also gives small performance variances on both tasks than the state-of-the-arts.

	\subsection{Ablation Study}
	% No Adversarial learning, Only Supervised, No MMD, One training stage
	% Performance by changing the weight for MMD loss
	
	\begin{figure}
		\begin{subfigure}{\linewidth}
			\includegraphics[width=\linewidth]{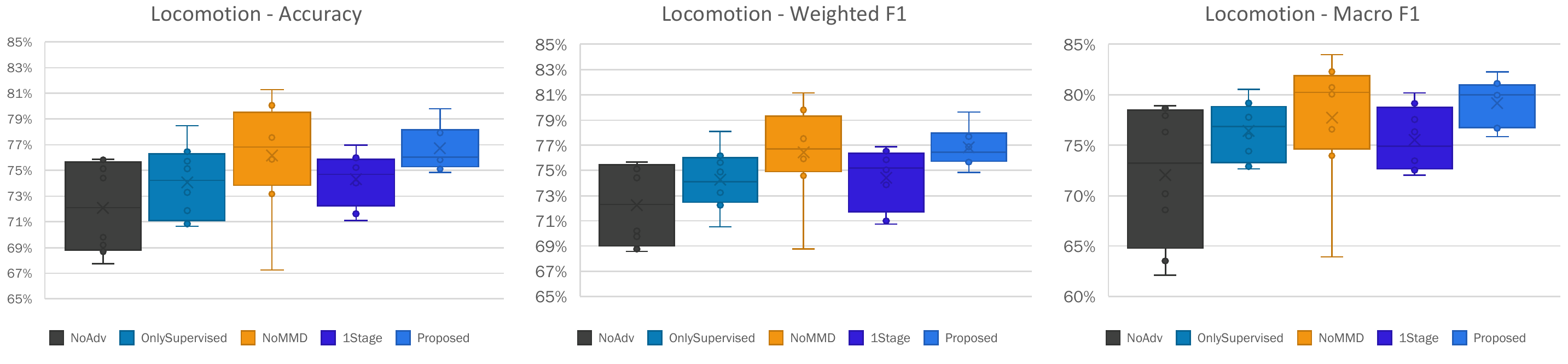}
			\caption{}
			\label{fig:ablationlocomotion}
		\end{subfigure}%
		\hfill
		\begin{subfigure}{\linewidth}
			\includegraphics[width=\linewidth]{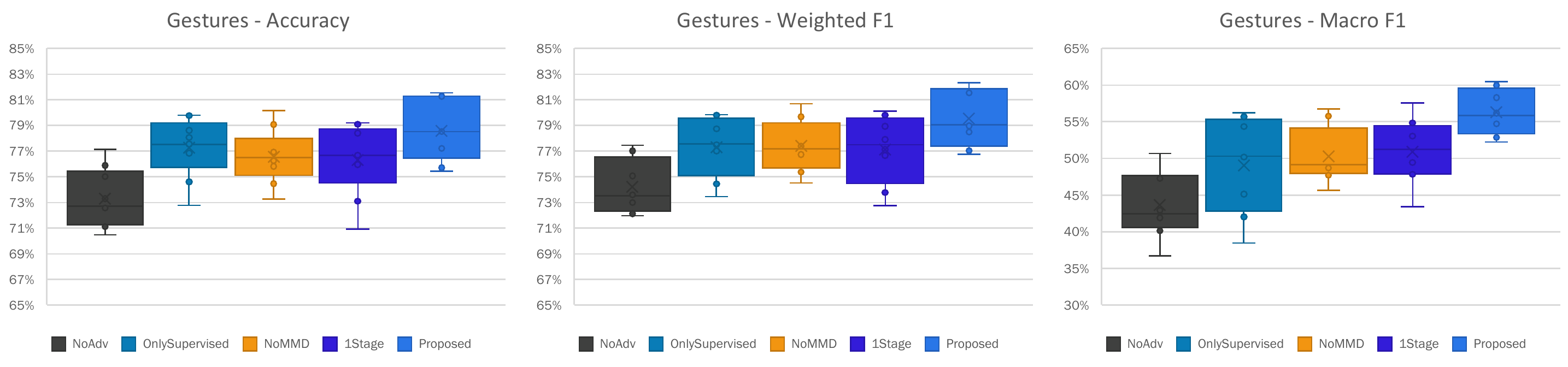}
			\caption{}
			\label{fig:ablationgestures}
		\end{subfigure}
		\hfill
		\begin{subfigure}{\linewidth}
			\includegraphics[width=\linewidth]{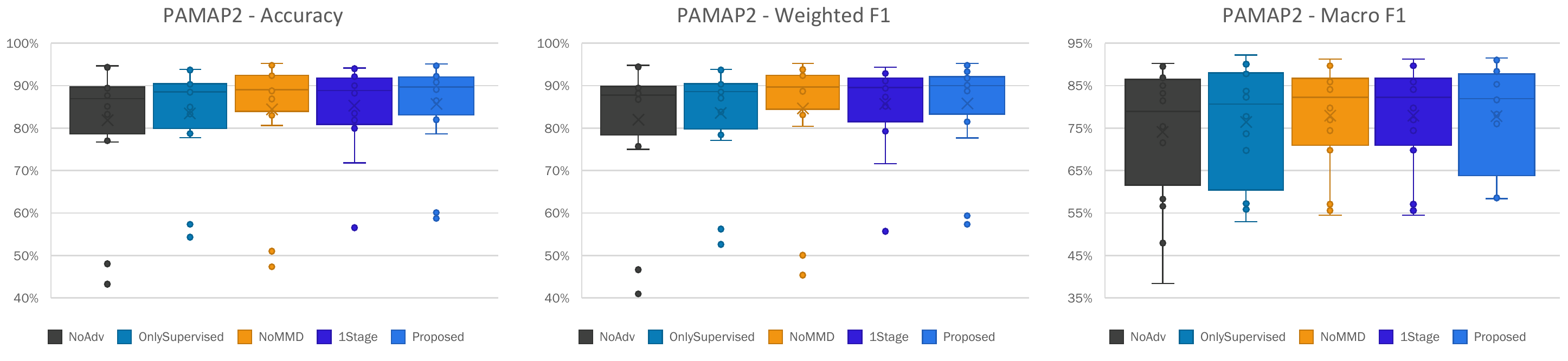}
			\caption{}
			\label{fig:ablationPAMAP2}
		\end{subfigure}
		\caption{Box-and-whisker plots of performance of the proposed method for ablation study on OPPORTUNITY (a) for the mode of locomotion recognition task, (b) for the gesture recognition, (c) and PAMAP dataset.}
		\label{fig:ablationstudy}
	\end{figure}

	To evaluate the effectiveness of the proposed adversarial feature extraction network architecture with MMD regularization, we conducted the ablation study. %In this section, we show the advantage of the proposed adversarial learning architecture and MMD regularization by changing the network architecture and loss functions.
	%In this section, we show the advantage of the proposed adversarial learning architecture and MMD regularization by changing the network architecture and loss functions, and show the effect of the hyperparameter $\lambda_{MMD}$ for regularization.
	
	\subsubsection{Effect of the proposed adversarial learning network architecture}
	%In this section, we show the advantage of the proposed adversarial learning architecture and MMD regularization by changing the network architecture and loss functions for optimization. 
	To show the effect of the proposed adversarial learning network architecture, we evaluate the proposed network architecture by changing the network architecture and loss functions for optimization.	
	\textbf{NoAdv} is the proposed activity classifier with the encoder-decoder architecture without the discriminator and adversarial learning to show the advantage of adversarial learning. \textbf{OnlySupervised} is the proposed adversarial feature extraction architecture and training procedure without the unsupervised learning for the target dataset to show the performance difference between with and without the unsupervised learning. \textbf{NoMMD} is to show the impact of the MK-MMD regularization by implementing the proposed adversarial learning without the MMD regularization. We denote \textbf{1Stage} as the proposed network architecture trained by only one stage training procedure, unlike the proposed method trained by three stages training procedure. 
	
	\cref{tab:ablationstudy} shows the evaluation results of the ablation study for the proposed adversarial learning network architecture with MMD regularization on the Opportunity and PAMAP2 dataset. The proposed method compared to other variants of the architecture showed better results in terms of $acc$, $F_w$, and $F_m$, indicating that the proposed adversarial learning enhances the generalization ability and improves the performance of the activity recognition. \textit{OnlySupervised} provided a better performance on the PAMAP2 dataset than on the Opportunity dataset. The Opportunity dataset contains only four subjects whereas the PAMAP dataset contains 8 subjects. It means that training the proposed model with more subjects can generalize the extracted features better. 
	
	Additionally, we analyze the evaluation results of the ablation study in the box-and-whisker plots, as shown in \cref{fig:ablationstudy}. The box-and-whisker plots of performance of the proposed method on the Opportunity dataset for the mode of locomotion recognition task in \cref{fig:ablationlocomotion} show that the proposed model provides the best performance with the smallest variance in terms of all three metrics, even though the highest performance values by \textit{NoMMD} is higher than the highest values by the proposed model. The box-and-whisker plots on the Opportunity dataset for the gesture recognition in \cref{fig:ablationgestures} also show that the proposed model outperforms other variants for the ablation study. The box-and-whisker plots of the performance on the PAMAP2 dataset show that the average performances look similar, but the outlier minimum performance of the proposed model is better than the one of other variants. In conclusion, the evaluation results for more difficult classification tasks showed that the performance of the proposed method provides better than the one of the variants clearly.
	
	\subsubsection{Effect of the hyperparameter for the MMD regularization}
	
	\label{subsubsec:hyperparameter}
	
	\begin{figure}
		\centering
		\includegraphics[width=\linewidth]{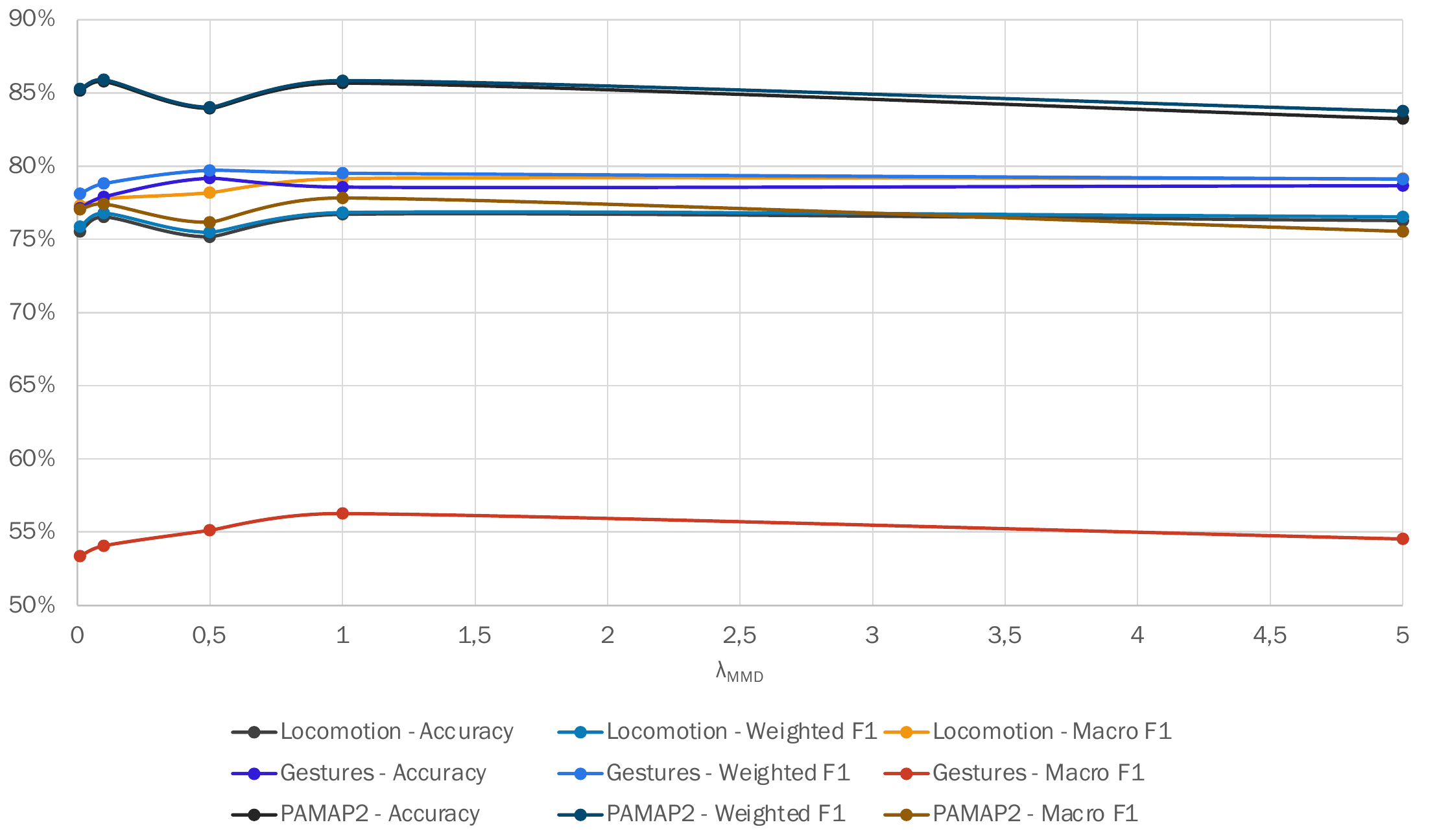}
		\caption{Performance on OPPORTUNITY and PAMAP2 datasets by changing hyperparameter $\lambda_{MMD}$.}
		\label{fig:hyperparameter_MMD}
	\end{figure}
	
	We perform experiments to better understand how much the hyperparameter $\lambda_{MMD}$ for the MMD regularization impacts the performance of the proposed model. We have conducted experiments with the hyperparameter $\lambda_{MMD}$ varying from 0.01 to 5.0 on the Opportunity dataset for two tasks and the PAMAP2 dataset. The bigger $\lambda_{MMD}$ means the optimization for the objective loss \cref{eq:objective} focuses more on the MMD regularization term \cref{eq:overallMMD} and the small $\lambda_{MMD}$ means the optimization for \cref{eq:objective} focuses more on the adversarial learning or activity classification.
	\cref{fig:hyperparameter_MMD} shows the effect of hyperparameter $\lambda_{MMD}$ on the performance in terms of $acc$, $F_w$, and $F_m$. Almost in all cases, the proposed method with $\lambda_{MMD}=1$ outperformed that with other hyperparameters, but the differences of the performances by different hyperparameter values were small. Thus, the hyperparameter value does not significantly affect the performance of the proposed model if the hyperparameter values are changed in the range of the scales.
	
\section{Conclusion} \label{sec:conclusion}
	
	In this study, we have proposed an adversarial feature extraction method with MMD regularization. The main idea of the proposed model is to learn a feature representation by joint optimization an encoder-decoder network structure regularized by the MK-MMD distance, an activity classifier, and a subject discriminator in an adversarial training manner. The encoder-decoder network structure for the feature extractor and reconstructor is designed to learn an embedding feature space to preserve the characteristics of the original signal, and the adversarial learning procedure between the feature extractor and subject discriminator learns the distributions of multiple domain data and extracts subject-invariant generalized features for activity recognition. The MK-MMD based regularization method helps enhance the generalization ability by aligning the feature distributions among different subject domains. The experimental results indicated that the proposed method can capture the important information for activity recognition and generalize the feature among different subject domains. Additionally, the proposed method outperformed the state-of-the-art methods and the ablation study showed that the advantage of the proposed model and objective loss function.
	
	For future work, we plan to extend the proposed method to the sensor-invariant feature extraction for human activity recognition. We found that the datasets we have tested used different types and numbers of sensors on different positions of the body. Recognizing the importance of each sensor or generalizing the features from different sensors can optimize their number and placement on the body and enable learning across datasets, domains and modalities.
	We look forward to reducing the labor-intensive data collection and annotation processes by utilizing the various datasets for different tasks.
	
% \section*{Acknowledgment}
% The preferred spelling of the word ``acknowledgment'' in America is without 
% an ``e'' after the ``g''. Avoid the stilted expression ``one of us (R. B. 
% G.) thanks $\ldots$''. Instead, try ``R. B. G. thanks$\ldots$''. Put sponsor 
% acknowledgments in the unnumbered footnote on the first page.

\bibliographystyle{IEEEtran}
\bibliography{ref}

\begin{thebibliography}{10}
\providecommand{\url}[1]{#1}
\csname url@samestyle\endcsname
\providecommand{\newblock}{\relax}
\providecommand{\bibinfo}[2]{#2}
\providecommand{\BIBentrySTDinterwordspacing}{\spaceskip=0pt\relax}
\providecommand{\BIBentryALTinterwordstretchfactor}{4}
\providecommand{\BIBentryALTinterwordspacing}{\spaceskip=\fontdimen2\font plus
\BIBentryALTinterwordstretchfactor\fontdimen3\font minus
  \fontdimen4\font\relax}
\providecommand{\BIBforeignlanguage}[2]{{%
\expandafter\ifx\csname l@#1\endcsname\relax
\typeout{** WARNING: IEEEtran.bst: No hyphenation pattern has been}%
\typeout{** loaded for the language `#1'. Using the pattern for}%
\typeout{** the default language instead.}%
\else
\language=\csname l@#1\endcsname
\fi
#2}}
\providecommand{\BIBdecl}{\relax}
\BIBdecl

\bibitem{lara2012survey}
O.~D. Lara and M.~A. Labrador, ``A survey on human activity recognition using
  wearable sensors,'' \emph{IEEE communications surveys \& tutorials}, vol.~15,
  no.~3, pp. 1192--1209, 2012.

\bibitem{wang2019deep}
J.~Wang, Y.~Chen, S.~Hao, X.~Peng, and L.~Hu, ``Deep learning for sensor-based
  activity recognition: A survey,'' \emph{Pattern Recognition Letters}, vol.
  119, pp. 3--11, 2019.

\bibitem{yang2015deep}
J.~Yang, M.~N. Nguyen, P.~P. San, X.~L. Li, and S.~Krishnaswamy, ``Deep
  convolutional neural networks on multichannel time series for human activity
  recognition,'' in \emph{Twenty-fourth international joint conference on
  artificial intelligence}, 2015.

\bibitem{ordonez2016deep}
F.~J. Ord{\'o}{\~n}ez and D.~Roggen, ``Deep convolutional and lstm recurrent
  neural networks for multimodal wearable activity recognition,''
  \emph{Sensors}, vol.~16, no.~1, p. 115, 2016.

\bibitem{mahmud2020human}
S.~Mahmud, M.~Tonmoy, K.~K. Bhaumik, A.~Rahman, M.~A. Amin, M.~Shoyaib,
  M.~A.~H. Khan, and A.~A. Ali, ``Human activity recognition from wearable
  sensor data using self-attention,'' \emph{arXiv preprint arXiv:2003.09018},
  2020.

\bibitem{cutting1977recognizing}
J.~E. Cutting and L.~T. Kozlowski, ``Recognizing friends by their walk: Gait
  perception without familiarity cues,'' \emph{Bulletin of the psychonomic
  society}, vol.~9, no.~5, pp. 353--356, 1977.

\bibitem{7966182}
M.~S. Singh, V.~Pondenkandath, B.~Zhou, P.~Lukowicz, and M.~Liwickit,
  ``Transforming sensor data to the image domain for deep learning — an
  application to footstep detection,'' in \emph{2017 International Joint
  Conference on Neural Networks (IJCNN)}, 2017, pp. 2665--2672.

\bibitem{saputri2014user}
T.~R.~D. Saputri, A.~M. Khan, and S.-W. Lee, ``User-independent activity
  recognition via three-stage ga-based feature selection,'' \emph{International
  Journal of Distributed Sensor Networks}, vol.~10, no.~3, p. 706287, 2014.

\bibitem{7317739}
J.-H. Hong, J.~Ramos, and A.~K. Dey, ``Toward personalized activity recognition
  systems with a semipopulation approach,'' \emph{IEEE Transactions on
  Human-Machine Systems}, vol.~46, no.~1, pp. 101--112, 2016.

\bibitem{chen2020metier}
L.~Chen, Y.~Zhang, and L.~Peng, ``Metier: A deep multi-task learning based
  activity and user recognition model using wearable sensors,''
  \emph{Proceedings of the ACM on Interactive, Mobile, Wearable and Ubiquitous
  Technologies}, vol.~4, no.~1, pp. 1--18, 2020.

\bibitem{sheng2020weakly}
T.~Sheng and M.~Huber, ``Weakly supervised multi-task representation learning
  for human activity analysis using wearables,'' \emph{Proceedings of the ACM
  on Interactive, Mobile, Wearable and Ubiquitous Technologies}, vol.~4, no.~2,
  pp. 1--18, 2020.

\bibitem{bai2020adversarial}
L.~Bai, L.~Yao, X.~Wang, S.~S. Kanhere, B.~Guo, and Z.~Yu, ``Adversarial
  multi-view networks for activity recognition,'' \emph{Proceedings of the ACM
  on Interactive, Mobile, Wearable and Ubiquitous Technologies}, vol.~4, no.~2,
  pp. 1--22, 2020.

\bibitem{arjovsky2017WGAN}
M.~Arjovsky, S.~Chintala, and L.~Bottou, ``Wasserstein generative adversarial
  networks,'' in \emph{International Conference on Machine Learning}, 2017, pp.
  214--223.

\bibitem{iwasawa2017privacy}
Y.~Iwasawa, K.~Nakayama, I.~Yairi, and Y.~Matsuo, ``Privacy issues regarding
  the application of dnns to activity-recognition using wearables and its
  countermeasures by use of adversarial training.'' in \emph{IJCAI}, 2017, pp.
  1930--1936.

\bibitem{gretton2006kernel}
A.~Gretton, K.~Borgwardt, M.~Rasch, B.~Sch{\"o}lkopf, and A.~Smola, ``A kernel
  method for the two-sample-problem,'' \emph{Advances in neural information
  processing systems}, vol.~19, pp. 513--520, 2006.

\bibitem{long2015learning}
M.~Long, Y.~Cao, J.~Wang, and M.~Jordan, ``Learning transferable features with
  deep adaptation networks,'' in \emph{International conference on machine
  learning}.\hskip 1em plus 0.5em minus 0.4em\relax PMLR, 2015, pp. 97--105.

\bibitem{zeng2017semi}
M.~Zeng, T.~Yu, X.~Wang, L.~T. Nguyen, O.~J. Mengshoel, and I.~Lane,
  ``Semi-supervised convolutional neural networks for human activity
  recognition,'' in \emph{2017 IEEE International Conference on Big Data (Big
  Data)}.\hskip 1em plus 0.5em minus 0.4em\relax IEEE, 2017, pp. 522--529.

\bibitem{varamin2018deep}
A.~A. Varamin, E.~Abbasnejad, Q.~Shi, D.~C. Ranasinghe, and H.~Rezatofighi,
  ``Deep auto-set: A deep auto-encoder-set network for activity recognition
  using wearables,'' in \emph{Proceedings of the 15th EAI International
  Conference on Mobile and Ubiquitous Systems: Computing, Networking and
  Services}, 2018, pp. 246--253.

\bibitem{chavarriaga2013opportunity}
R.~Chavarriaga, H.~Sagha, A.~Calatroni, S.~T. Digumarti, G.~Tr{\"o}ster,
  J.~d.~R. Mill{\'a}n, and D.~Roggen, ``The opportunity challenge: A benchmark
  database for on-body sensor-based activity recognition,'' \emph{Pattern
  Recognition Letters}, vol.~34, no.~15, pp. 2033--2042, 2013.

\bibitem{reiss2012introducing}
A.~Reiss and D.~Stricker, ``Introducing a new benchmarked dataset for activity
  monitoring,'' in \emph{2012 16th international symposium on wearable
  computers}.\hskip 1em plus 0.5em minus 0.4em\relax IEEE, 2012, pp. 108--109.

\bibitem{banos2014mhealthdroid}
O.~Banos, R.~Garcia, J.~A. Holgado-Terriza, M.~Damas, H.~Pomares, I.~Rojas,
  A.~Saez, and C.~Villalonga, ``mhealthdroid: a novel framework for agile
  development of mobile health applications,'' in \emph{International workshop
  on ambient assisted living}.\hskip 1em plus 0.5em minus 0.4em\relax Springer,
  2014, pp. 91--98.

\bibitem{bello2021MoCapaci}
H.~Bello, B.~Zhou, S.~Suh, and P.~Lukowicz, ``Mocapaci: Posture and gesture
  detection in loose garments using textile cables as capacitive antennas,'' in
  \emph{2021 International Symposium on Wearable Computers}, 2021, pp. 78--83.

\bibitem{cook2013transfer}
D.~Cook, K.~D. Feuz, and N.~C. Krishnan, ``Transfer learning for activity
  recognition: A survey,'' \emph{Knowledge and information systems}, vol.~36,
  no.~3, pp. 537--556, 2013.

\bibitem{deng2014cross}
W.-Y. Deng, Q.-H. Zheng, and Z.-M. Wang, ``Cross-person activity recognition
  using reduced kernel extreme learning machine,'' \emph{Neural Networks},
  vol.~53, pp. 1--7, 2014.

\bibitem{zhao2011cross}
Z.~Zhao, Y.~Chen, J.~Liu, Z.~Shen, and M.~Liu, ``Cross-people mobile-phone
  based activity recognition,'' in \emph{Twenty-second international joint
  conference on artificial intelligence}, 2011.

\bibitem{wang2018stratified}
J.~Wang, Y.~Chen, L.~Hu, X.~Peng, and S.~Y. Philip, ``Stratified transfer
  learning for cross-domain activity recognition,'' in \emph{2018 IEEE
  international conference on pervasive computing and communications
  (PerCom)}.\hskip 1em plus 0.5em minus 0.4em\relax IEEE, 2018, pp. 1--10.

\bibitem{khan2018scaling}
M.~A. A.~H. Khan, N.~Roy, and A.~Misra, ``Scaling human activity recognition
  via deep learning-based domain adaptation,'' in \emph{2018 IEEE international
  conference on pervasive computing and communications (PerCom)}.\hskip 1em
  plus 0.5em minus 0.4em\relax IEEE, 2018, pp. 1--9.

\bibitem{faridee2019augtoact}
A.~Z.~M. Faridee, M.~A. A.~H. Khan, N.~Pathak, and N.~Roy, ``Augtoact: Scaling
  complex human activity recognition with few labels,'' in \emph{Proceedings of
  the 16th EAI International Conference on Mobile and Ubiquitous Systems:
  Computing, Networking and Services}, 2019, pp. 162--171.

\bibitem{akbari2019transferring}
A.~Akbari and R.~Jafari, ``Transferring activity recognition models for new
  wearable sensors with deep generative domain adaptation,'' in
  \emph{Proceedings of the 18th International Conference on Information
  Processing in Sensor Networks}, 2019, pp. 85--96.

\bibitem{zhao2020local}
J.~Zhao, F.~Deng, H.~He, and J.~Chen, ``Local domain adaptation for
  cross-domain activity recognition,'' \emph{IEEE Transactions on Human-Machine
  Systems}, vol.~51, no.~1, pp. 12--21, 2020.

\bibitem{wang2018deep}
J.~Wang, V.~W. Zheng, Y.~Chen, and M.~Huang, ``Deep transfer learning for
  cross-domain activity recognition,'' in \emph{proceedings of the 3rd
  International Conference on Crowd Science and Engineering}, 2018, pp. 1--8.

\bibitem{jeyakumar2019sensehar}
J.~V. Jeyakumar, L.~Lai, N.~Suda, and M.~Srivastava, ``Sensehar: a robust
  virtual activity sensor for smartphones and wearables,'' in \emph{Proceedings
  of the 17th Conference on Embedded Networked Sensor Systems}, 2019, pp.
  15--28.

\bibitem{goodfellow2014generative}
I.~Goodfellow, J.~Pouget-Abadie, M.~Mirza, B.~Xu, D.~Warde-Farley, S.~Ozair,
  A.~Courville, and Y.~Bengio, ``Generative adversarial nets,'' in
  \emph{Advances in neural information processing systems}, 2014, pp.
  2672--2680.

\bibitem{ioffe2015batch}
S.~Ioffe and C.~Szegedy, ``Batch normalization: Accelerating deep network
  training by reducing internal covariate shift,'' in \emph{International
  conference on machine learning}.\hskip 1em plus 0.5em minus 0.4em\relax
  Lille, France: PMLR, 2015, pp. 448--456.

\bibitem{hochreiter1997long}
S.~Hochreiter and J.~Schmidhuber, ``Long short-term memory,'' \emph{Neural
  computation}, vol.~9, no.~8, pp. 1735--1780, 1997.

\bibitem{he2016deep}
K.~He, X.~Zhang, S.~Ren, and J.~Sun, ``Deep residual learning for image
  recognition,'' in \emph{Proceedings of the IEEE conference on computer vision
  and pattern recognition}, 2016, pp. 770--778.

\bibitem{vaswani2017attention}
A.~Vaswani, N.~Shazeer, N.~Parmar, J.~Uszkoreit, L.~Jones, A.~N. Gomez,
  {\L}.~Kaiser, and I.~Polosukhin, ``Attention is all you need,'' in
  \emph{Advances in neural information processing systems}, 2017, pp.
  5998--6008.

\bibitem{nguyen2015recognizing}
L.~T. Nguyen, M.~Zeng, P.~Tague, and J.~Zhang, ``Recognizing new activities
  with limited training data,'' in \emph{Proceedings of the 2015 ACM
  International Symposium on Wearable Computers}, 2015, pp. 67--74.

\bibitem{ma2019attnsense}
H.~Ma, W.~Li, X.~Zhang, S.~Gao, and S.~Lu, ``Attnsense: Multi-level attention
  mechanism for multimodal human activity recognition.'' in \emph{IJCAI}, 2019,
  pp. 3109--3115.

\bibitem{kingma2014adam}
D.~P. Kingma and J.~Ba, ``Adam: A method for stochastic optimization,''
  \emph{arXiv preprint arXiv:1412.6980}, 2014.

\end{thebibliography}

\end{document}